\documentclass[twocolumn,12pt]{aastex62}
\pdfoutput=1

\shorttitle{DRIFTING PULSATION STRUCTURE IN THE 2017 SEPTEMBER 10 FLARE}
\shortauthors{Karlick\'y et al.}
\begin{document}

\title{DRIFTING PULSATION STRUCTURE AT THE VERY BEGINNING OF\\ THE 2017 SEPTEMBER 10 LIMB FLARE}

\author[0000-0002-3963-8701]{Marian Karlick\'{y}}
\affiliation{Astronomical Institute of the Czech Academy of Sciences,
Fri\v{c}ova 298, Ond\v{r}ejov, 251 65, Czech Republic}

\author[0000-0002-0660-3350]{Bin Chen}
\affiliation{Center for Solar-Terrestrial Research, New Jersey Institute of
Technology, 323 M L King Jr Boulevard,  Newark, NJ 07102-1982, USA}

\author[0000-0003-2520-8396]{Dale E. Gary}
\affiliation{Center for Solar-Terrestrial Research, New Jersey Institute of
Technology, 323 M L King Jr Blvd,
   Newark, NJ 07102-1982, USA}

\author[0000-0001-9559-4136]{Jana Ka\v{s}parov\'{a}}
\affiliation{Astronomical Institute of the Czech Academy of Sciences,
Fri\v{c}ova 298, Ond\v{r}ejov, 251 65, Czech Republic}

\author[0000-0003-3128-8396]{Jan Ryb\'ak}
\affiliation{Astronomical Institute, Slovak Academy of Sciences, Tatransk\'{a}
Lomnica, Slovakia}


\begin{abstract}
Drifting pulsation structures (DPSs) are important radio fine structures
usually observed at the beginning of eruptive solar flares. It has been
suggested that DPSs carry important information on the energy release processes
in solar flares. We study DPS observed in an X8.2-class flare on 2017 September
10 in the context of spatial and spectral diagnostics provided by microwave,
EUV, and X-ray observations. We describe DPS and its substructures that were
observed for the first time. We use a new wavelet technique to reveal
characteristic periods in DPS and their frequency bands.  Comparing the periods
of pulsations found in this DPS with those in previous DPSs we found new very
short periods in the 0.09--0.15 s range. We present Expanded Owens Valley Solar Array images and spectra of
microwave sources observed during the DPS. This DPS at its very beginning has
pulsations in two frequency bands (1000--1300 MHz and 1600--1800 MHz) which are
interconnected by fast drifting bursts. We show that these double-band
pulsations started just at the moment when the ejected filament splits apart in
a tearing motion at the location where a signature of the flare current sheet
later appeared. Using the standard flare model and previous observations of
DPSs, we interpret these double-band pulsations as a radio signature of
superthermal electrons trapped in the rising magnetic rope and flare arcade at
the moment when the flare magnetic reconnection starts. The results are
discussed in a scenario with the plasmoid in the rising magnetic rope.
\end{abstract}


\section{Introduction}

Solar flares classified in soft X-rays (\textit{GOES}) as X-flares belong to the
strongest type of flares, thus they are a topic of great interest. The 2017
September 10 X8.3 flare, which was partially occulted by the west limb and
peaked at around 16:00 UT, was the second largest solar flare in Solar Cycle
24. It has been studied from various points of view in more than 30 papers to
date.

For example, \cite{2018ApJ...853L..18Y} show that the flare was associated with
a flux-rope eruption followed by the plasma inflow with the formation of a
current sheet and coronal mass ejection. Analyzing the spectra of
high-temperature \ion{Fe}{24} lines observed by the Extreme-Ultraviolet Imaging
Spectrometer (EIS) in the impulsive phase of this flare,
\cite{2018ApJ...864...63P} found that the lines are very broad. They interpret
this broadening as caused by the presence of a non-Maxwelian electron
distribution. They also estimated the plasma density of the hot plasma
component in the flare arcade at 16:59 UT as $(0.9-2.0) \times 10^{11}$
cm$^{-3}$. Furthermore, based on Expanded Owens Valley Solar Array (EOVSA) and
the \textit{Reuven Ramaty High Energy Solar Spectroscopic Imager} (\textit{RHESSI)} observations,
\cite{2018ApJ...863...83G} show that the microwave and HXR sources arise from a
common nonthermal electron population, although the microwave sources occupy a
much larger area. \cite{2018ApJ...854..122W} presented spectroscopic
observations of the current sheet in this flare and estimated the emission
measure and plasma density in the current sheet as 10$^{30}$ cm$^{-5}$ and
10$^{10}$ cm$^{-3}$, respectively. \cite{2018ApJ...855...74L} studied the
coronal cavity around the erupting filament. The magnetic morphology and
dynamics of the accompanying coronal mass ejection were studied by
\cite{2018ApJ...868..107V}. \cite{2018ApJ...865L...7O} analyzed the gamma-ray
emission of this flare, and suggested three phases in acceleration of protons
lasting more than 12 hr. \cite{2018ApJ...863L..39G} reported the ground-level
enhancement event associated with a coronal mass ejection whose initial
acceleration ($\sim$ 9.1 km s$^{-2}$) and initial speed ($\sim$ 4300 km
s$^{-1}$) were among the highest observed by the LASCO coronagraph.

Although the drifting pulsation structure (DPS) was recorded sometimes in older
radio spectra \citep[see, e.g. the Figure 24 in ][]{1994A&AS..104..145I}, DPS as a
distinct radio burst was established after its physical interpretation in the
paper by \cite{2000A&A...360..715K}. In this paper, based on the observations
made by \cite{1998ApJ...499..934O} and using the numerical modeling of the
magnetic reconnection it was proposed that DPS is a plasma emission signature
of the plasmoid formed in the impulsive phase of solar flares.

DPSs occur quite frequently at the beginning of the eruptive solar flares,
mainly in the 1000--2000 MHz frequency range \citep{2015ApJ...799..126N}. They
consist of many relatively narrowband pulses that show a frequency drift to
lower or sometimes higher frequencies as a whole.  Their physical origin can
be explained within the standard CSHKP flare model
\citep{1964NASSP..50..451C,1966Natur.211..695S,1974SoPh...34..323H,1976SoPh...50...85K}.
Before the flare, a magnetic rope (i.e. current-carrying loop) is formed, owing
to shear and vortex plasma flows at the photospheric level. At its bottom part,
a cold and dense plasma condenses into the filament. Then this magnetic rope
together with the filament somehow becomes unstable, through proposed
mechanisms such as the torus or kink instabilities
\citep{2006PhRvL..96y5002K,2010SoPh..266...91K} and/or a decrease of the
stabilizing magnetic force of the above-lying magnetic field lines
\citep{2011ApJ...730...57A}. This loss of stability leads to upward motion of
the magnetic rope. Then, the current sheet is formed below this rising magnetic
rope. When this current sheet becomes sufficiently narrow, magnetic
reconnection starts. Electrons are accelerated at or near the reconnection
X-point. They move toward the rising filament and also downward to the flare
arcade. Some of these electrons are then trapped in the upper part of the
rising filament or/and in secondary ropes (formed because of the plasmoid
instability in the current sheet; \citealt{2007PhPl...14j0703L}). These trapped
superthermal electrons in the magnetic rope or in secondary ropes can
simultaneously generate the X-ray emission by bremsstrahlung
\citep{1998ApJ...499..934O}, microwaves by the gyrosynchrotron mechanism
\citep{2010SoPh..266...91K}, and DPS due to plasma emission. The sources of
these emissions associated with the magnetic ropes are designated in
observations as plasmoids. (Note that the term "plasmoid" is also used in the
plasmoid instability theory in a somewhat different sense.) The pulses in DPS
are owing to a quasi-periodic acceleration of electrons at the X-point of the
magnetic reconnection. During the upward motion of the magnetic rope, the
plasma density inside the plasmoid located in the magnetic rope decreases,
thus the emission frequency decreases and DPS as a whole drifts to lower
frequencies.

After the initial paper by \cite{2000A&A...360..715K}, DPSs were studied in several later
papers
\citep{2002A&A...388..363K,2004A&A...419..365K,2007A&A...464..735K,2008SoPh..253..173B,2008A&A...477..649B},
where observational and model details of DPS were presented. Nevertheless,
there are still many open questions regarding DPSs, especially on their
substructures.

In this paper, using high-time resolution radio spectra
\citep{2008SoPh..253...95J}, we present an example of DPS with clearly seen
substructures and very short period pulsations. This DPS at its beginning
consists of pulsations in two frequency bands (1000--1300 MHz and 1600--1800 MHz)
which are interconnected by fast drifting bursts. Furthermore, DPS reveals
substructures as e.g. narrowband continua and positively fast drifting pulses.
To provide context, we show the associated spatial and spectral radio observations obtained by EOVSA, as well as EUV and X-ray
observations. (Note that such spatial radio observations associated with DPS
are presented for the first time.) Finally, we interpret these double-band
pulsations at the start of the DPS in the framework of the eruptive solar flare model,
together with results of previous observations and numerical simulations of DPS. Our
motivation is to describe the DPS, its substructures and associate phenomena
in detail and account for the most remarkable substructure of DPS: two pulsation
bands at the beginning of DPS.

The paper is organized as follows. In Section 2 we describe
DPS, its substructures, analysis of pulsations and associated observations in
radio (EOVSA), EUV and X-rays. The interpretation is in Section 3. Discussions
and conclusions are in Section 4.

\section{Observations}

For the analysis of the 2017 September 10 flare, SOL2017-09-10, we use observations
in radio from the Ond\v{r}ejov radiospectrograph with the time resolution 0.01
s \citep{1993SoPh..147..203J,2008SoPh..253...95J} and EOVSA
\citep{2018ApJ...863...83G}, in X-rays from \textit{FERMI} Gamma-Ray Burst Monitor (GBM)
\citep{2009ApJ...702..791M}) and \textit{RHESSI} \citep{2002SoPh..210....3L}, and in EUV
from \textit{SDO}/AIA~\citep{2012SoPh..275...17L}.

\subsection{Spectral observations of DPS and its wavelet analysis}

\begin{figure*}
\begin{center}
\includegraphics[width=16cm]{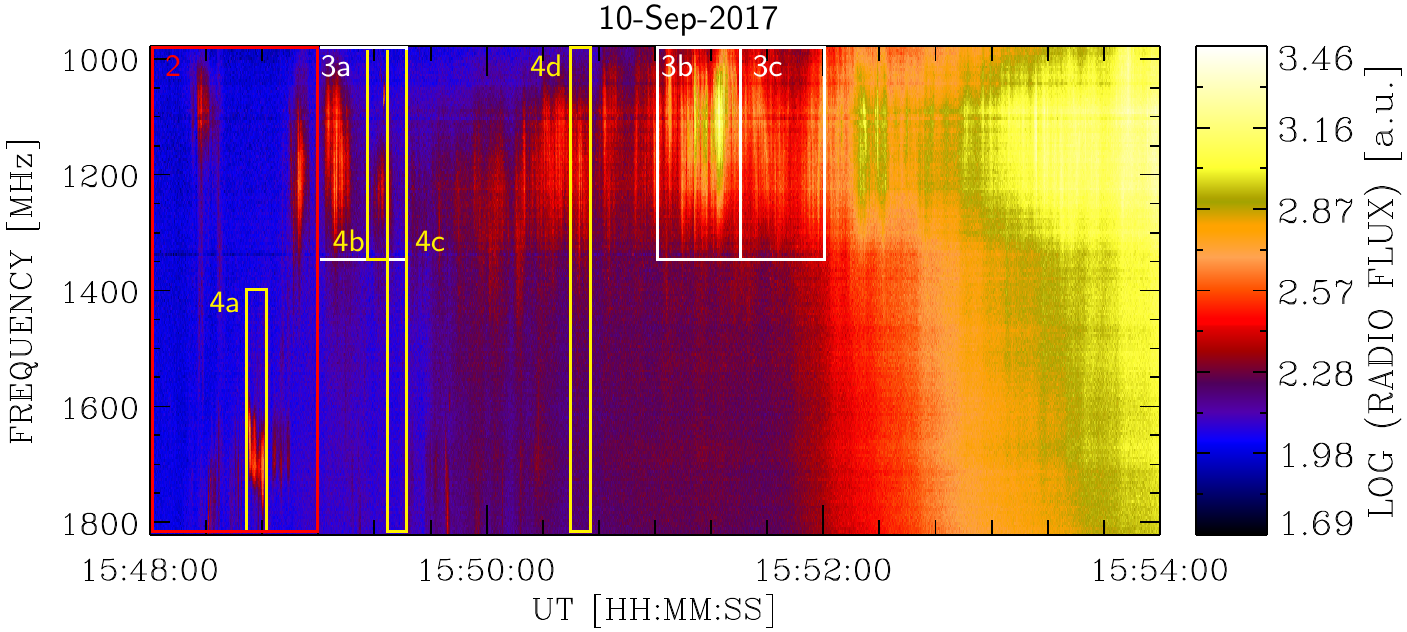}
\end{center}
  \caption{Overview radio dynamic spectrum in the 1000-1800 MHz range observed during the 2017 September 10 flare by the Ond\v{r}ejov radiospectrograph \citep{2008SoPh..253...95J}
  at 15:48-15:54 UT. The color boxes outline regions seen at higher zoom level in other figures, and the numbers
  and letters indicate the corresponding figure and panel numbers.}
  \label{fig_glob}
\end{figure*}

The 1000--1800 MHz radio spectrum observed at the very beginning of the 2017
September 10 flare is shown in Figure~\ref{fig_glob}. Because the emission in
this band became so bright as to saturate the Ond\v{r}ejov spectrograph after
15:54 UT, we limit our discussion to the very early phase of the flare
(15:48--15:54 UT), which ends 5-6 minutes before the peak of the impulsive phase
($\sim$16:00~UT). The radio emission as a whole during this early time is a
classic example of DPS
\citep{2000A&A...360..715K,2004A&A...419..365K,2008SoPh..253..173B}. It starts
at 15:48:12 UT in two frequency bands at about 1000--1300 MHz and 1600--1800
MHz. Then from 15:50:00 to 15:51:30 UT a series of pulsations in the range
1000--1300 MHz drifts toward lower frequencies with a frequency drift rate of about
$-$1.6 MHz s$^{-1}$. After 15:51:30 UT the pulsations remain within the same
frequency range of 1000--1300 MHz superimposed on a strongly brightening
continuum covering the entire 1000--1800 MHz band. A detail of the initial
radio emission is shown in Figure~\ref{fig_initial}. It is important to see
that at 15:48:17--15:48:25 UT the two frequency bands (1000--1300 MHz and
1600--1800 MHz) are interconnected with fast drifting bursts.

On higher frequencies (2--4.5 GHz), the spectrum shows the broadband continuum
starting at about 15:48:30 UT; see
\url{http://www.asu.cas.cz/~radio/radio/rt4/2017/A1709102.gif}. On the other
hand, at frequencies below 1~GHz, the e-Callisto (BIR) 200--400 MHz spectrum
shows type III bursts and continua, and the e-Callisto (BLENSW) 20--80 MHz
spectrum type III and II bursts
(\url{http://soleil.i4ds.ch/solarradio/callistoQuicklooks/}).

The radio spectrum in Figure~\ref{fig_glob} was observed with a time resolution
of 10~ms. This offers a unique opportunity to search for substructures in this
DPS. Examples of substructures at higher time resolution are shown in
Figures~\ref{fig_30s} and ~\ref{fig_6s} (see boxes in Figure~\ref{fig_glob},
which place these panels in the context of the overall emission).
Figure~\ref{fig_30s} shows pulsations in the 1000--1300 MHz range in 30 s time
intervals. As seen here, not only does the bandwidth of some pulsations vary,
but in some cases their upper- or lower-frequency boundaries drift to higher or
lower frequencies. Examples of the substructures in selected 6 s intervals are
shown in Figure~\ref{fig_6s}. It is interesting that most pulsations show a
positive frequency drift on this time scale, see (Figures~\ref{fig_6s}(b) and ~\ref{fig_6s}(d)).
For example, the burst in the 1150--1300 MHz range at
15:49:22.6--15:49:22.7 UT has a drift rate of about 1.5 GHz s$^{-1}$ and the burst in
the 1100--1300 MHz range at 15:50:33.75--15:50:33.9 UT a drift rate of about 1.3 GHz
s$^{-1}$. On the other hand, in pulsations in the 1600--1800 MHz frequency
range presented in Figure~\ref{fig_6s}(a), no measurable frequency drift was
found. However, the whole series of pulsations (15:48:35--15:48:41 UT) slowly
drifts toward higher frequencies with the frequency drift $\sim$10 MHz s$^{-1}$.
Figure~\ref{fig_6s}(c) shows a double stripe burst with stripe frequencies of
1080 and 1120 MHz at 15:49:25 UT. See also that this burst is associated with
the pulses in the 1700--1900 MHz frequency range. We pay attention to this
double stripe burst because it is very bright (700 SFU) and it resembles the
very bright zebra-like burst shown by \cite{2014ApJ...790..151T}. A variety of
different substructures (pulsations, narrowband continua, drifting pulses,
dot-like bursts) indicates complex electron distribution functions in the DPS
source. The drifting pulses could be connected with the beam-like distribution,
while the pulsations and narrowband continua with the beam-like or/and with
loss-cone distributions.

\subsubsection{Wavelet analysis of pulsations in DPS}

As shown in numerical simulations by \cite{2000A&A...360..715K}, the magnetic
reconnection has a quasi-periodic character. Thus, in the magnetic reconnection
electrons are accelerated quasi-periodically and generate a quasi-periodic
radio emission. To obtain the time scales in the acceleration process
in the flare magnetic reconnection in more detail, we analyzed the observed
pulsations in the DPS using the wavelet method recently described in
\cite{2017SoPh..292....1K}. We adopt the Morlet wavelet analysis method with
$\omega = 6$. Only periods with 99$\%$ significance are presented. This method
not only identifies the oscillation periods of the pulsations, but also
provides the corresponding phase information.  For each time interval in the
radio dynamic spectrum that contains DPS features, we calculate a histogram of
pulsation periods. Selecting the range of periods at which the histogram peaks,
we use the corresponding phase information to identify which features display
those dominant periods. The ``phase dynamic spectrum'' is superposed on the
corresponding radio dynamic spectrum, shown in Figures ~\ref{fig_hist_spe1} and
\ref{fig_hist_spe3} as pink colors of varying brightness, with black
corresponding to zero phase. The time interval between neighboring black lines
in these maps indicates the chosen period.

Two examples of our wavelet analysis are shown in Figures~\ref{fig_hist_spe1}
and \ref{fig_hist_spe3}. Histograms in these figures show periods from 0.1 up
to 2 s. In the lower panel of Figure~\ref{fig_hist_spe1}, we show the
spectrum with the period of of 0.7--1.1 s which is an example of typical
periods found in the analyzed DPS. However, the lower panel of
Figure~\ref{fig_hist_spe3} shows the oscillations with much shorter periods of
0.09--0.15 s at 15:48:40--15:48:41 UT in the 1650--1750 MHz frequency band. Some
indications of this period is also at 15:48:35.3--15:48:36:3 UT in the 1600--1700
MHz band. The periods around 1 s and longer are common
\citep[see e.g.][]{2000A&A...360..715K,2017SoPh..292...94K,2018SoPh..293...62K}, but the
periods around 0.1 s in the DPS substructure are new and  have not been
reported before.

\subsection{Associated UV, EOVSA, and X-Ray observations}

In Figure~\ref{fig_171} we show the \textit{SDO}/AIA 171 \AA\ observations at the time
of double-band pulsations at the beginning of DPS. It is remarkable that at
this time the ejected filament is tearing (at about 15:48:20 UT) in the region
(the arrow in Figure~\ref{fig_171}(b)), where at later times a signature of the flare
current sheet appeared. At this moment the EOVSA multifrequency radio sources
fall along the entire erupting structure, with the lower-frequency contours
lying close to the tearing region (Figure~\ref{fig_4849}(a)). About 40 s later,
the lower-frequency source ($<$5.4 GHz) starts to bifurcate into two parts
(Figure ~\ref{fig_4849}(b)).

Later, the upper part of the ruptured filament moves upward, see the 3.41 GHz
red source in the green box in Figure~\ref{fig_eovsa}(a). Figure~\ref{fig_eovsa}(a)
shows positions of the 3.4--18 GHz radio sources in four regions designated
A--D, while Figure~\ref{fig_eovsa}(b) shows the light curves on 3.41 GHz from
these four different regions, marked with a box with the corresponding letter
and color in Figure~\ref{fig_eovsa}(a). Comparing these four sources with the
magnetic rope designated by the red dashed line presented in Figure 3 in
\cite{2018ApJ...853L..18Y}, source A corresponds to a part of the rapidly
rising flux rope observed only on the lowest EOVSA frequencies, source B
corresponds to the flare arcade observed in the full EOVSA frequency range (in
the cyan box), and sources C and D on either side of the flare arcade coincide
roughly with the footpoints of the magnetic rope. After 15:56 UT the green
light curve (corresponding to the green box A) is no longer due to the flux-rope
source, but is dominated by the emission from the upward-growing arcade
source, which encroaches into the green box.

At flare stage, $\sim$15:52~UT, we also derived the spatially resolved
brightness temperature spectra that correspond to each of the four EOVSA microwave
sources (Figure~\ref{fig_eovsa}(c)). The spectrum from source B at the top of the
flare arcade is well measured, and is consistent with nonthermal emission, as
its peak brightness temperature is above 60 MK. The other sources (A, C, and D) are also possibly nonthermal, inferred from their peak brightness temperature of $\gtrsim$ 30 MK. The nonthermal nature of these sources, however is less certain, due to having less statistically significant frequency points (filled circles; open circles are data points dominated by
noise).

In Figure~\ref{fig_X}, we compare the EOVSA 3.41 GHz light curves
(Figure~\ref{fig_eovsa}(b)), \textit{FERMI}/GBM 50--100 keV hard emission and the
1000--1300 MHz radio flux corresponding to the main frequency band of DPS. From
a global point of view, the DPS 1000--1300 MHz radio flux, 50--100 keV hard
X-ray and 3.41 GHz light curves are correlated. As concerns individual peaks,
it appears that the blue and green light curves of sources D and A (southern
and that above the flare arcade) show some similarity to DPS and hard X-rays
around 15:51:30 UT. However, in such a comparison it is necessary to take into
account that the hard X-rays, EOVSA radio and DPS are generated by quite
different emission mechanisms: by bremsstrahlung, gyrosychrotron, and plasma
emission mechanisms, respectively. While the hard X-ray emission depends on the
electron distribution of the superthermal electrons and the background plasma
density, the EOVSA radio emission depends not only on the superthermal electron
distribution, but on the magnetic field in the radio source. On the other hand,
the brightness of the plasma emission depends on the derivative of the electron
distribution in momentum space. At the start of DPS, there is no spatial HXR
information available. (The first HXR image could be reconstructed no earlier
than at 15:53 UT.) Figure~\ref{fig_X_source} shows positions of the X-ray
sources at this time. No further X-ray sources out of the field of view of this
figure were found. Relative to the EOVSA radio sources, the X-ray sources are
located only at comparatively low altitudes.

\section{Interpretation}\label{sec:interpret}

The most remarkable substructure is the double-band pulsations at the very
beginning of DPS. Unfortunately, we do not have spatial observations at the DPS
frequencies. Moreover, determination of the plasma density from EUV data in the
assumed DPS source at this very early flare phase is not possible, due to the
lack of appropriate data. Therefore, we interpret these double-band pulsations
indirectly using the standard CSHKP flare model, previous observations of DPSs
and results of numerical modeling of the magnetic reconnection. We also add
further arguments in favor of our interpretation.

As already mentioned, comparing the positions of four (A, B, C, and D) EOVSA
sources (Figure~\ref{fig_eovsa}(a)) with the magnetic rope designated by the red
dashed line in Figure 3 of \cite{2018ApJ...853L..18Y}, we can see
that the source A is in the upper part of the magnetic rope, source B in the
upper part of the flare arcade, and sources C and D near footpoints of the
magnetic rope. According to the standard CSHKP flare model, the X-point of the
flare magnetic reconnection, where particles are likely accelerated, is located
between source A and B. From numerical modeling of the magnetic reconnection it
is known that the accelerated electrons propagate from the X-point of the
magnetic reconnection in the direction of the reconnection plasma outflows \citep{2003PhPl...10.3554R,2006JGRA..11110212P,2008ApJ...674.1211K},
i.e., in our case, upward in the solar atmosphere to source A, and downward
to source B. The source in the upper part of the rising magnetic rope, i.e.,
source A, is referred to as a plasmoid. The plasmoid can be observed in X-rays as
generated by the superthermal electrons by bremsstrahlung
\citep{1998ApJ...499..934O} and/or in radio, the present source A, or the 17
GHz plasmoid \citep{2010SoPh..266...71K} generated by the gyrosynchrotron
mechanism. In papers by \cite{2000A&A...360..715K,2008SoPh..253..173B}, it was
shown that the superthermal electrons in such a plasmoid can also generate DPS.

Therefore, we assume that source A is also the source of the present DPS.
Source A is the nonthermal source where there are the superthermal
electrons which could generate DPS. Furthermore, comparing sources in
Figure~\ref{fig_4849} at 15:48:24 and 15:49:08 UT with source A at 15:52:16
UT (Figure~\ref{fig_eovsa}), we can see that source A rises. It agrees with
the negative drift of DPS; see also numerical simulations in
\cite{2008A&A...477..649B}.

The magnetic reconnection below the rising filament starts when the current
sheet is sufficiently narrow. Right at the beginning of magnetic reconnection, the
distance between the magnetic rope and flare arcade is relatively small, therefore there
is also small difference in plasma densities. At this moment, the electrons
accelerated at the X-point of the magnetic reconnection move between the rising
magnetic rope and the flare arcade and bombard the flare arcade as well as the
magnetic rope. We think that this process exhibits in our observations as the
double-band pulsations interconnected by the fast drifting bursts; see
Figure~\ref{fig_initial}. Namely, the superthermal electrons trapped in the
magnetic rope and upper part of the flare arcade and along their trajectories
between these traps generate the observed double-band pulsations and fast
drifting bursts by the plasma emission mechanism.

This interpretation is supported by the observation of a tearing of the ejected
filament at the same time as that of the double-band pulsations, and at the
location where in the later time the flare current sheet was recognized, and a
bifurcation of the lower-frequency source ($<$5.4 GHz) 40 s later (Figure
~\ref{fig_4849}(b)). Moreover, DPS in the 1000--1300 MHz as a whole drifts
negatively, and the narrowband continuum in the 1600--1800 MHz range
(Figure~\ref{fig_6s}(a)), which belongs to the high-frequency part of the
double-band pulsations, drifts positively. It expresses the upward motion of
the rising magnetic rope to lower plasma densities and the increase of the
plasma density in the arcade, respectively, which is in agreement with our
interpretation.

As concerns sources C and D (Figure~\ref{fig_eovsa}), our interpretation is that these
sources are generated by the superthermal electrons propagating along the
magnetic rope from the acceleration site (presumably located at or closed to source B according to the CSHKP scenario) by the gyrosynchrotron mechanism.

\section{Discussion and conclusions}

In the paper, we present DPS with substructures (narrowband continua, drifting
pulses, dot-like bursts) detected in DPSs for the first time. These different
substructures indicate complex electron distribution functions in the DPS
source. The drifting pulses could be associated with a beam-like distribution,
whereas the pulsations and narrowband continua with a beam-like or/and
loss-cone-type distribution. Thus, in the magnetic field of the magnetic rope
these distributions can cause the plasmoid to be visible not only in the plasma
emissions as substructures of DPS, but also in the gyrosynchrotron emission at
frequencies above 3.41 GHz.

Using our new method based on the wavelet transform, we analyzed the periods in
the time-frequency domain of DPS in detail. In most parts of DPS, we found
periods around 1 s, which is in agreement with our previous studies. However,  in the
narrowband substructure at times between 15:48:35 and 15:48:41 UT, in the
1650--1750 MHz frequency band we detected new and very short periods in the
0.09--0.15 s range.

We presented positions of EOVSA sources observed in 3.4--12.4 GHz, their intensity evolutions and spectra
at times of DPS. As followed from comparison of four EOVSA sources with the
magnetic rope designated in Figure 3 in 
\cite{2018ApJ...853L..18Y}, the EOVSA plasmoid (source A) coincides with the erupting magnetic
rope. As shown in Figure~\ref{fig_eovsa}(c), the spectra of the four microwave
sources (as in Figure \ref{fig_eovsa}(a)) are likely nonthermal, with their peak
brightness temperature in the 30--60 MK range. This indicates a presence of
superthermal electrons in all of these sources. However, while the spectrum of
source B is broadband, the spectra of the rest of the sources are relatively
narrowband, and peak at lower frequencies. Assuming that the electron flux from
the reconnection region upward to the flux rope and downward to the arcade is
roughly the same then the question about the difference of spectra of the
microwave sources arises. We assume that it is not only due to higher magnetic
field strength in the arcade than that in the plasmoid, but also the arcade region
likely has additional processes of energy release and particle acceleration.

Based on the standard CSHKP flare model and the previous observations of DPSs,
we interpret that the EOVSA plasmoid (source A) is also the plasmoid where DPS is
generated; see Section~\ref{sec:interpret}. In this scenario, all periods found in DPS can be explained
by the periodicity of electron acceleration in the magnetic reconnection below
the magnetic rope as proposed by ~\cite{2000A&A...360..715K}. However, there
are other possibilities, e.g. that electron beam circulates in the
magnetic field of the plasmoid, generating the plasma emission visible only
from one side of the plasmoid, emitting the drifting pulses as in our case. In
such a model, the period is given by the length of trajectory around the
plasmoid divided by the beam velocity.

In our interpretation, the bulk of the nonthermal electrons are accelerated in
or near the magnetic reconnection site located between the rising plasmoid (or
the magnetic rope) and the flare arcade. Electrons trapped above the flare arcade produce the looptop HXR (blue contours in Figure~\ref{fig_X_source}) and microwave source (source B in Figure~\ref{fig_eovsa}). Those precipitated at the footpoints give rise to the HXR footpoint source (blue and green contours near the solar limb in Figure~\ref{fig_X_source}). Some of the accelerated electrons are trapped in the flux rope/plasmoid (EOVSA source A in our case), and generate the emission at frequencies above 3.41 GHz by the gyrosynchrotron emission mechanism and also DPS by the plasma emission mechanism. The accelerated electrons in the flux rope may further escape and propagate to the footpoints of the flux rope, which we adopt as one possible interpretation for the presence of the two EOVSA microwave side sources (sources C and D in Figure~\ref{fig_eovsa}), as well as the apparent similarity in their light curves with the central EOVSA microwave sources (sources B and A), the DPS, and X-ray emission during the initiation phase of the flare (prior to $\sim$15:54 UT; Figure~\ref{fig_X}). The observation of these side sources is somewhat similar to the microwave source observed by NoRH at 17 GHz that appeared to coincide with a hot magnetic rope (seen by AIA 131 \AA) during the preimpulsive stage of an eruptive 2012 July 19 flare \citep{2016ApJ...820L..29W}, although the peak brightness temperature of their 17 GHz microwave source was less than only 20 kK and spectral information was not available. 

\begin{figure*}
\begin{center}
  \includegraphics[width=16cm]{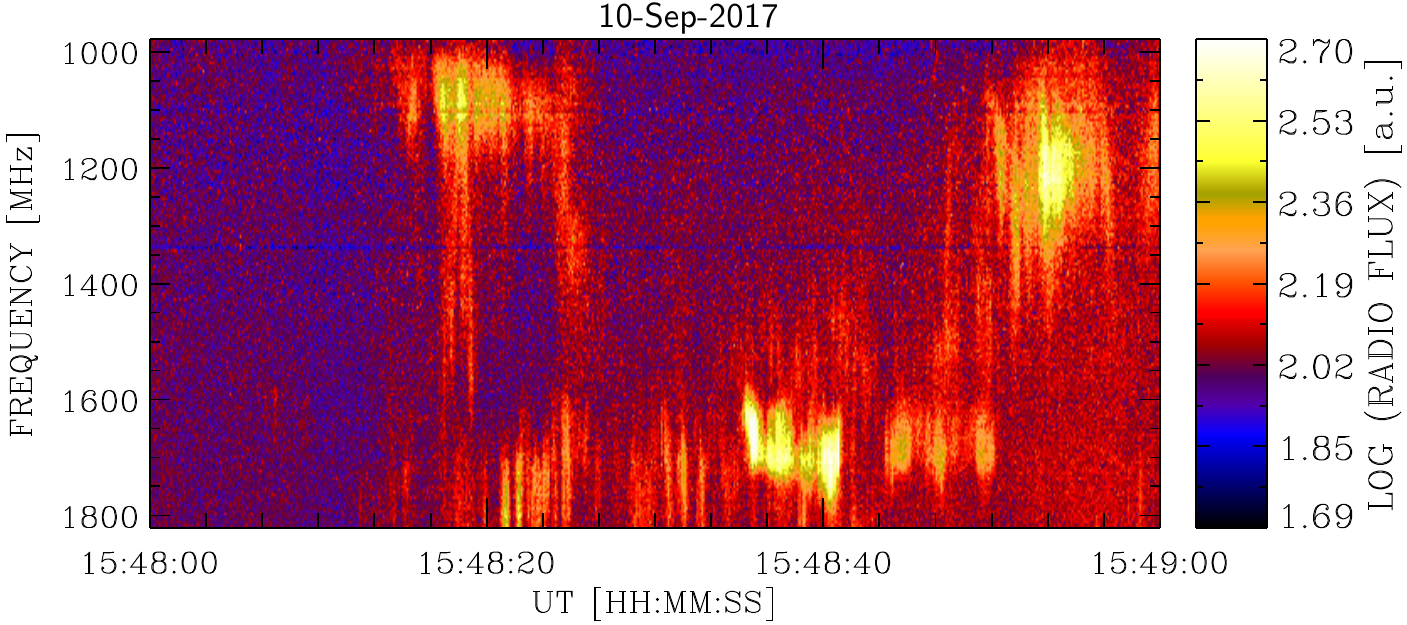}
\end{center}
  \caption{Detail of the radio dynamic spectrum in the 1000-1800 MHz range observed at the very beginning of the 2017 September 10 flare at 15:48-15:49 UT. The pulsations appear mainly in two frequency bands (1000--1300 MHz and 1600--1800 MHz), which are interconnected by fast drifting bursts.}
  \label{fig_initial}
\end{figure*}

\begin{figure}
\begin{center}
\includegraphics[width=10cm]{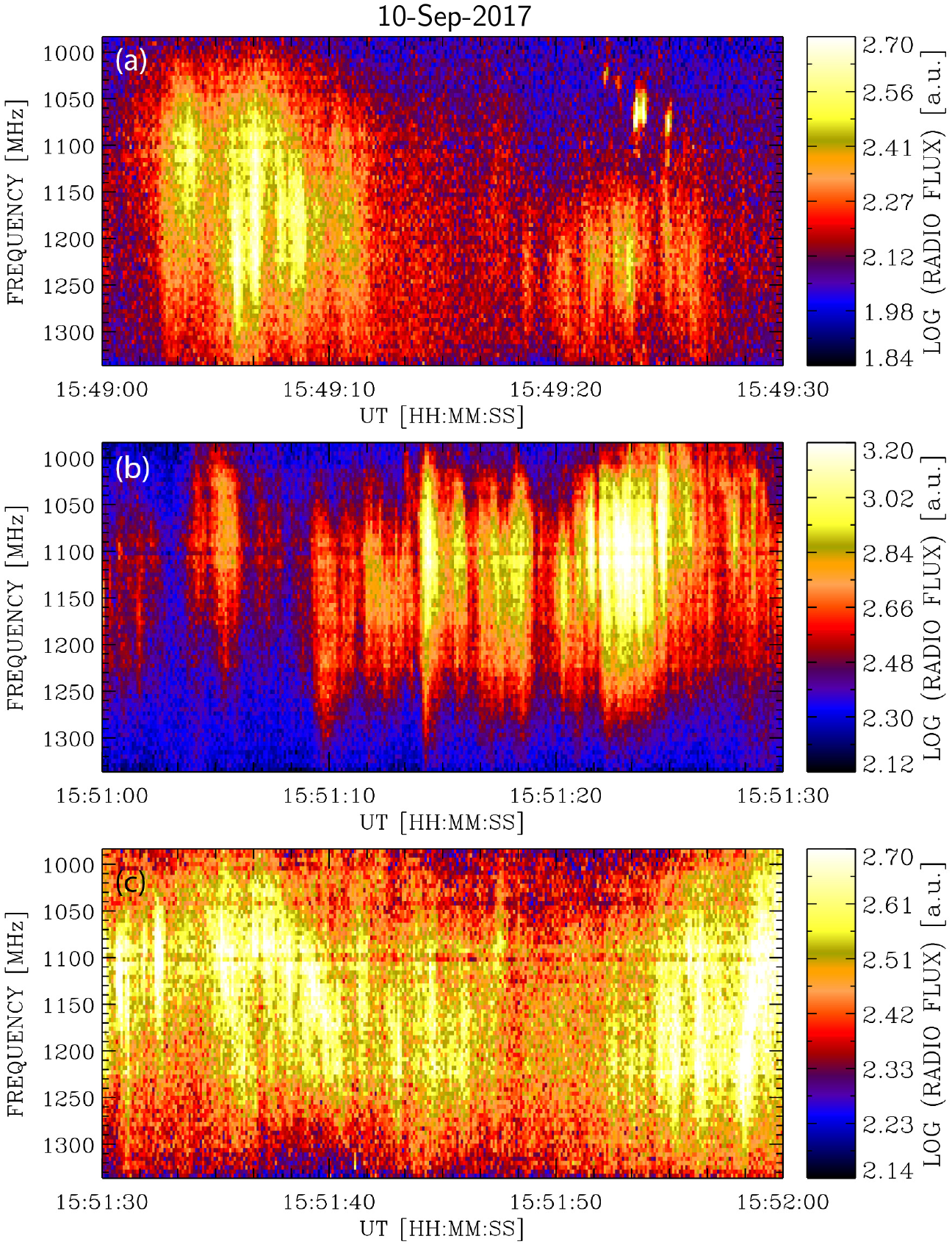}
\end{center}
  \caption{Three examples of the detailed 30 s radio dynamic spectra taken during the 15:48--15:54 UT time interval of the 2017 September 10 flare.
  See the white boxes labeled 3a, 3b, and 3c in Figure~\ref{fig_glob}.}
  \label{fig_30s}
\end{figure}

\begin{figure}
\begin{center}
\includegraphics[width=10cm]{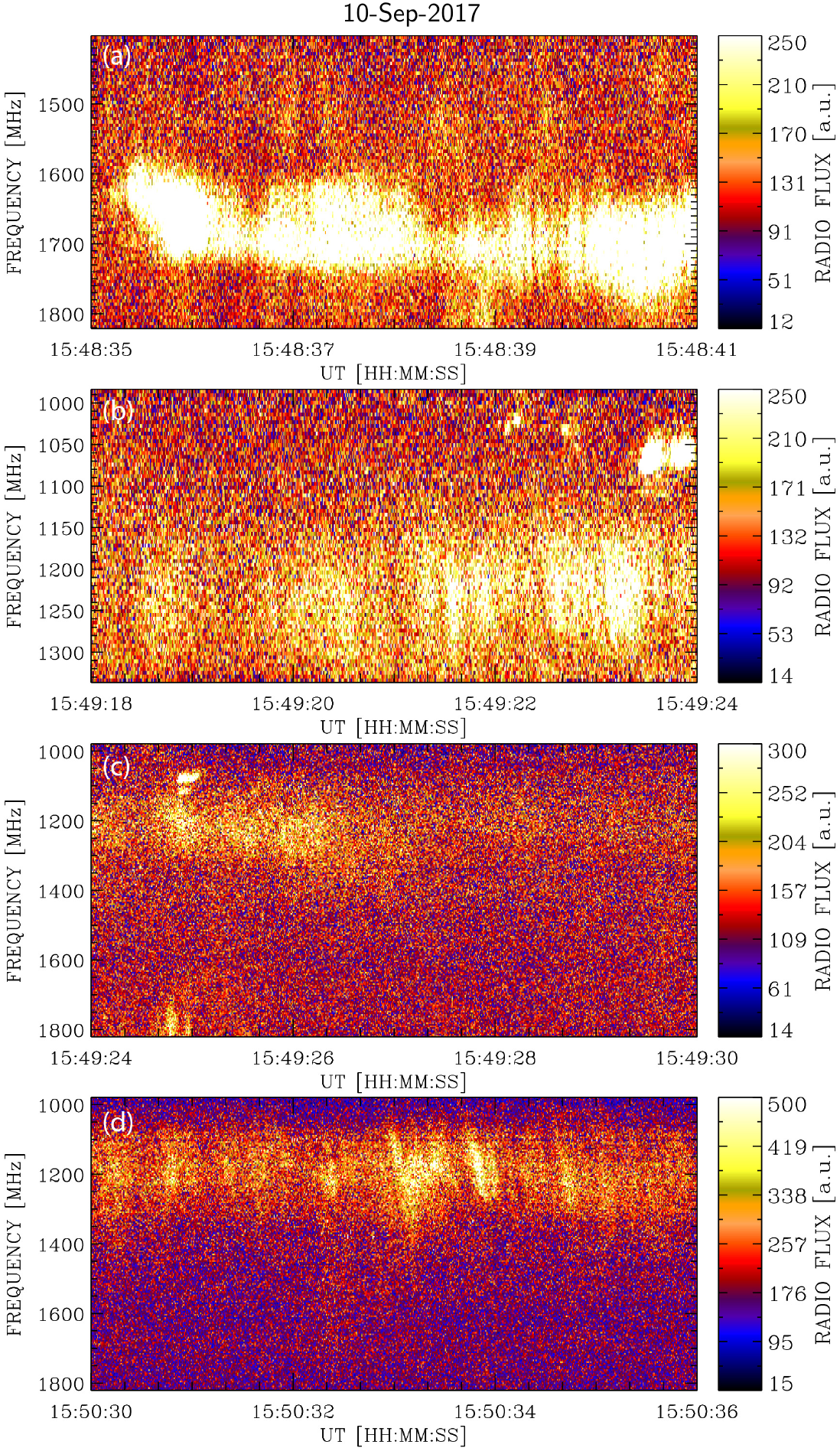}
\end{center}
\caption{Four examples of the detailed 6 s radio dynamic spectra taken during the 15:48--15:54 UT time interval of
  the 2017 September 10 flare. See the white boxes labeled 4a, 4b, 4c, and 4d in Figure~\ref{fig_glob}.}
  \label{fig_6s}
\end{figure}

\begin{figure}
\begin{center}
\includegraphics[width=13cm]{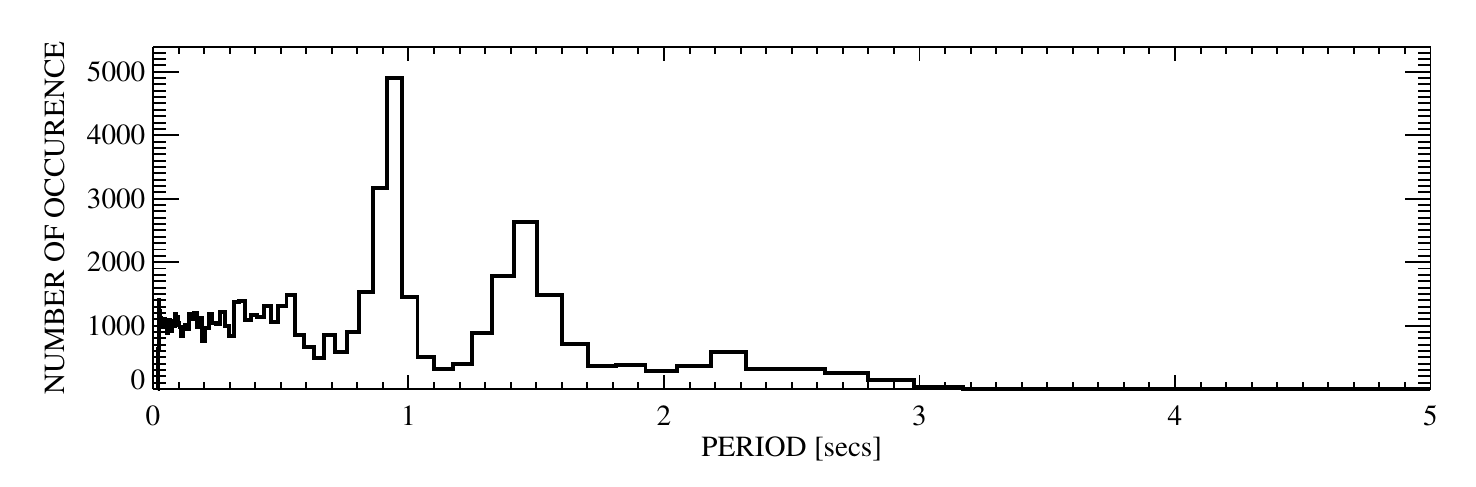}
\includegraphics[width=13cm]{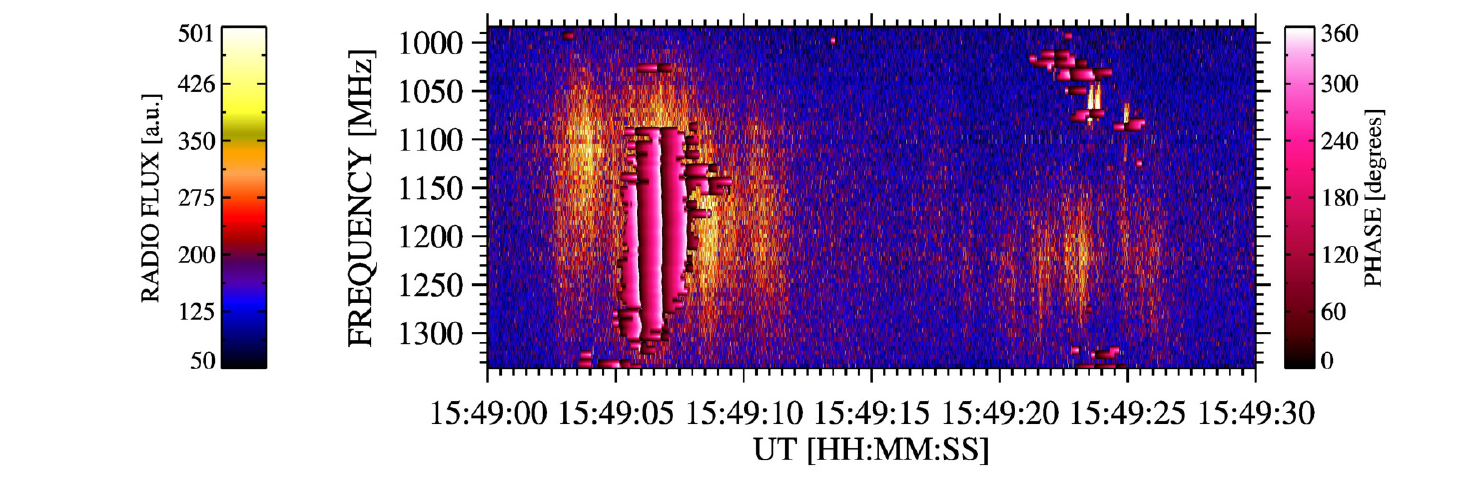}
\end{center}
  \caption{Histogram of periods in the spectrum observed at 15:49:00--15:49:30 UT (the spectrum in Figure~\ref{fig_30s}(a)) and the corresponding phase maps (pink areas with the black lines showing the zero phase of oscillations) overplotted on the radio dynamic spectrum for periods detected in the range of 0.7--1.1 s.}
  \label{fig_hist_spe1}
\end{figure}

\begin{figure}
\begin{center}
\includegraphics[width=13cm]{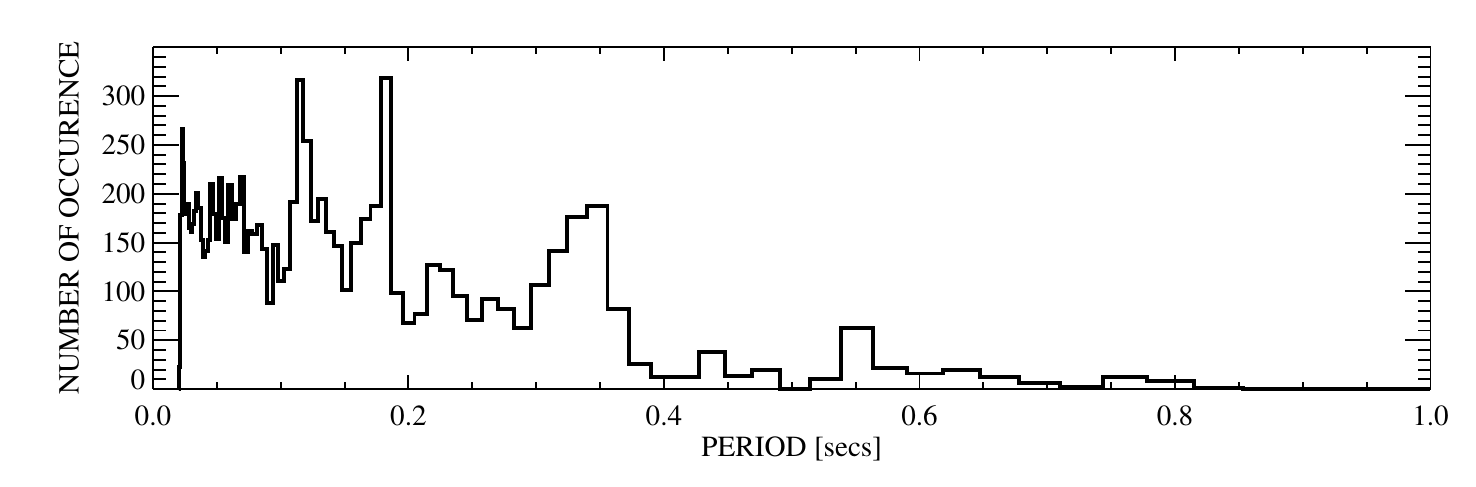}
\includegraphics[width=13cm]{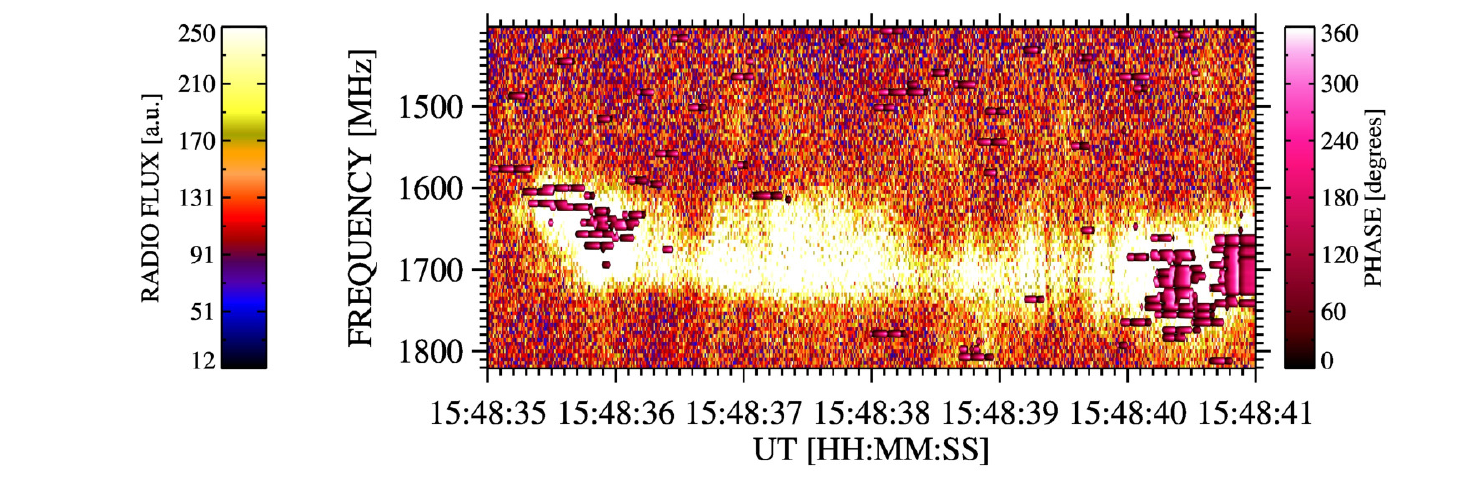}
\end{center}
  \caption{Histogram of periods in the spectrum observed at 15:48:35--15:48:41 UT (the spectrum in Figure~\ref{fig_6s}(a)) and the corresponding
phase maps (pink areas with the black lines showing the zero phase of oscillations)
overplotted on the radio dynamic
spectrum for periods detected in the range of 0.09--0.15 s.}
  \label{fig_hist_spe3}
\end{figure}

\begin{figure}
\begin{center}
\includegraphics[width=9cm]{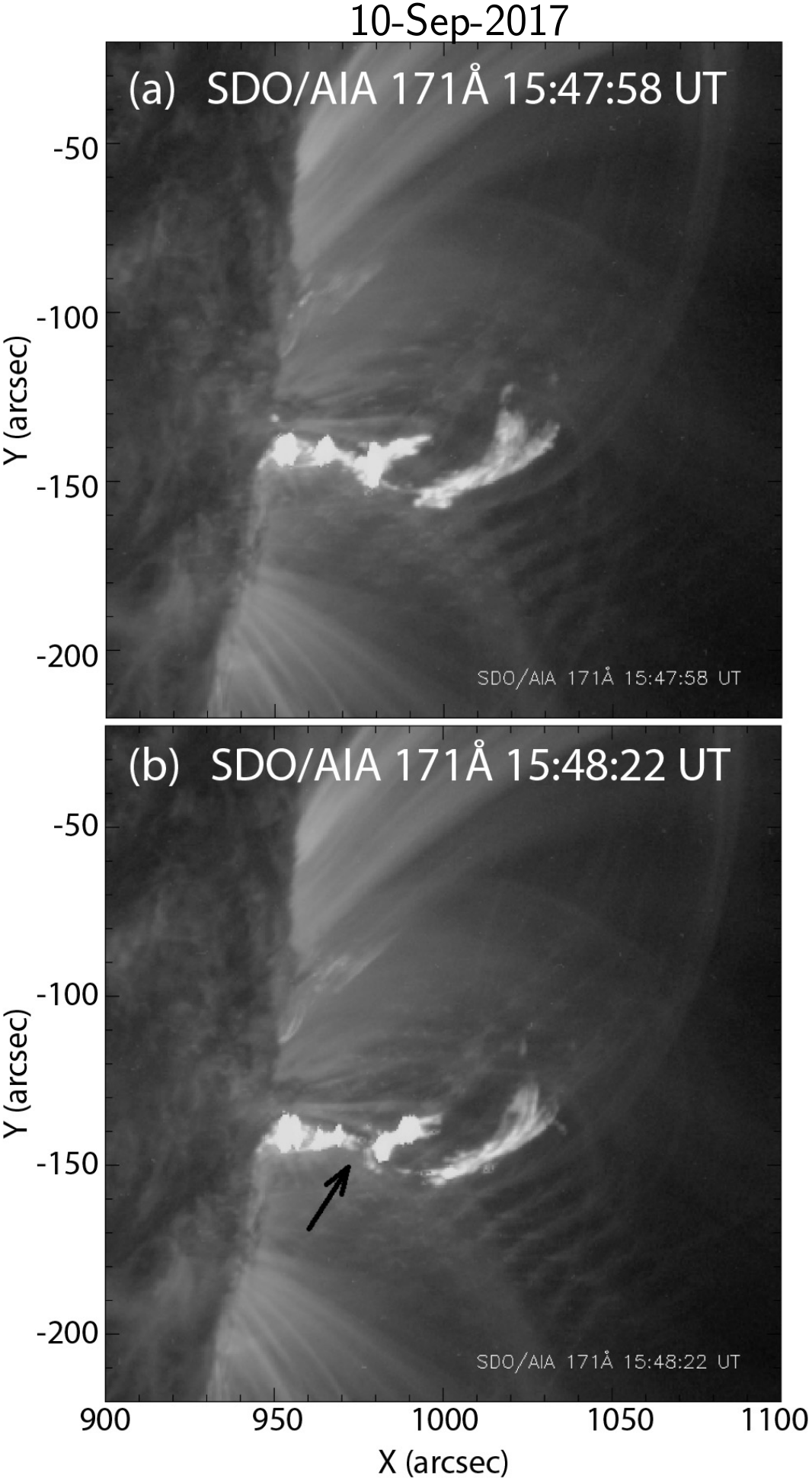}
\end{center}
  \caption{\textit{SDO}/AIA 171 \AA~~images at 15:47:58 and 15:48:22 UT during the 2017 September 10 flare, showing a tearing of the ejected filament,
  in the region marked by the black arrow, at 15:48:22 UT when on the radio spectrum the pulsations
  appeared in two frequency bands (Figure~\ref{fig_initial}).}
  \label{fig_171}
\end{figure}

\begin{figure*}
\begin{center}
\includegraphics[width=16cm]{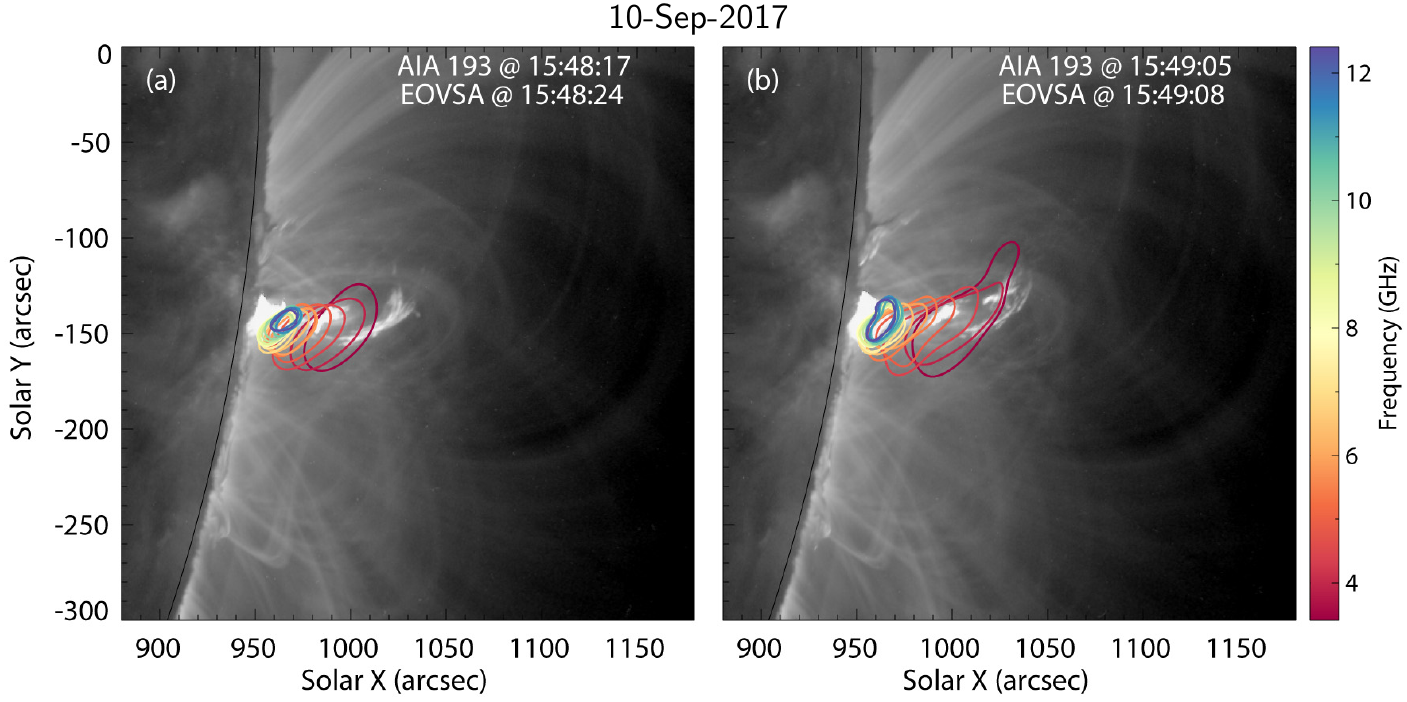}
\end{center}
  \caption{The EOVSA sources (contours) at 15:48:24 and 15:49:08 UT during the 2017 September 10 flare,
  in the region of a tearing of the ejected filament.}
  \label{fig_4849}
\end{figure*}

\begin{figure*}
\begin{center}
\includegraphics[width=1.0\textwidth]{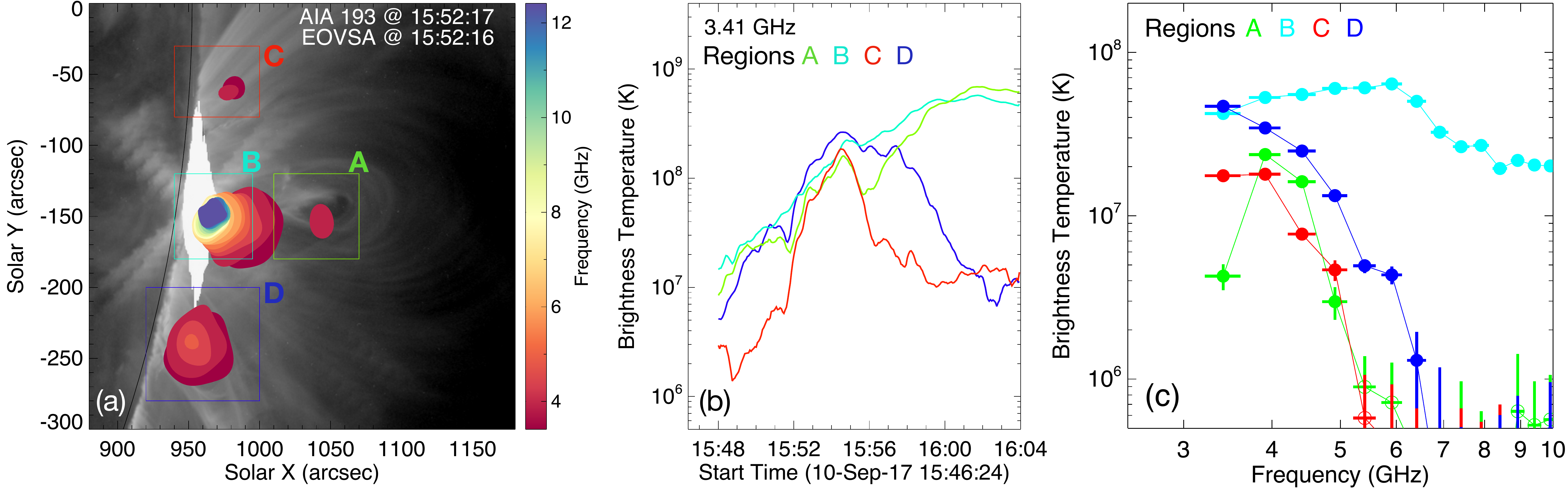}
\end{center}
  \caption{(a) EOVSA spectral imaging observation of the microwave sources in 3.4--12.4 GHz at 15:52:16 UT during the 2017 September 10 flare,
  with hues showing the sources at different frequencies (color bar). Background is \textit{SDO}/AIA 193~\AA~ image.
  (b) Time evolution of EOVSA 3.41 GHz maximum brightness temperature in four selected regions,
  marked with a box with the respective color in (a).
  (c) The corresponding spatially resolved brightness temperature spectra obtained near the center of each respective microwave sources.}
  \label{fig_eovsa}
\end{figure*}

\begin{figure}
\begin{center}
\includegraphics[width=10cm]{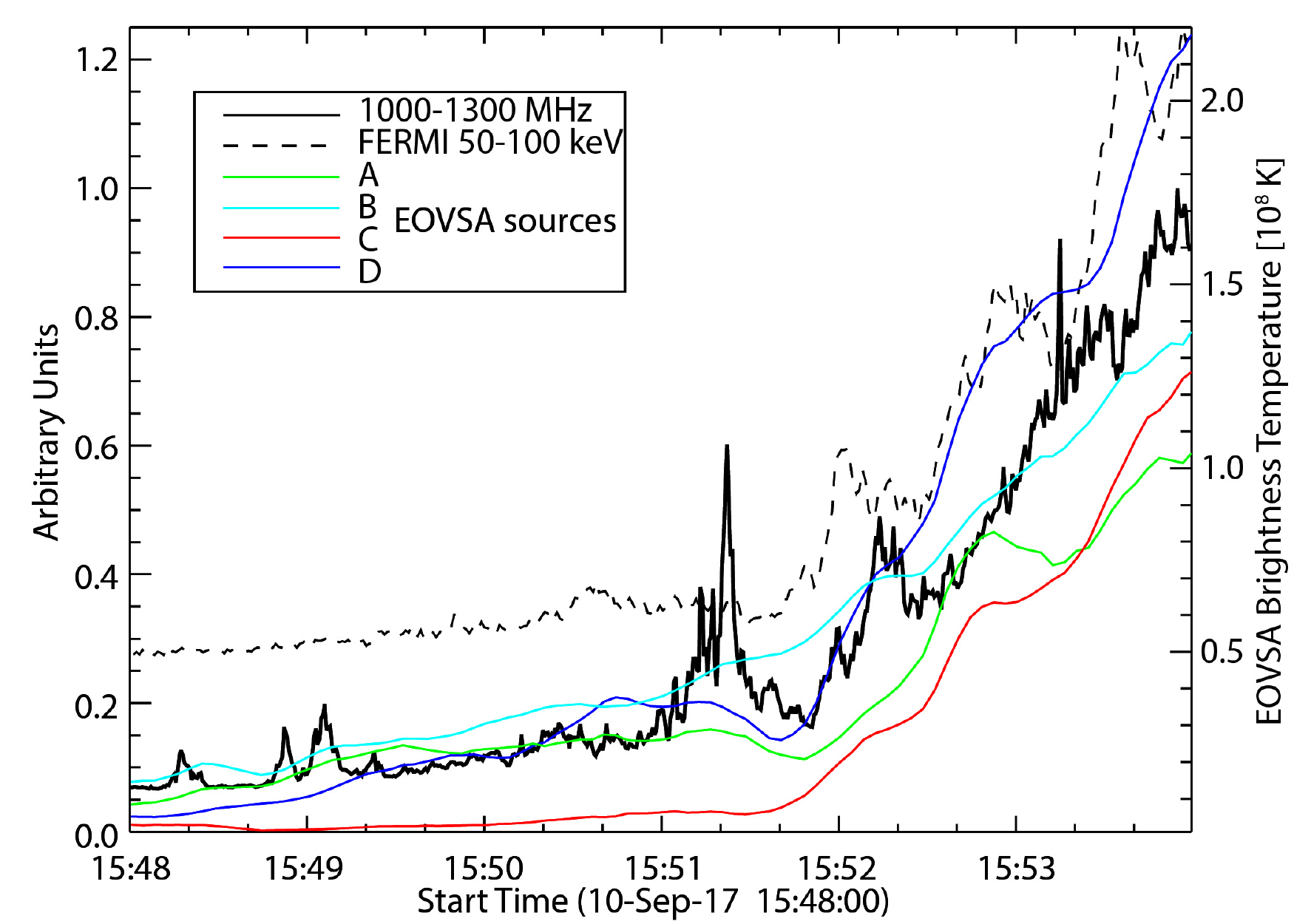}
\end{center}
  \caption{Radio flux in the 1000-1300 MHz range at 15:48--15:54 UT (black solid line) during the 2017 September 10 flare in comparison with the Fermi 50--100 keV X-ray flux (black dashed line)
  and the EOVSA light curves according to Figure~\ref{fig_eovsa}(b).}
  \label{fig_X}
\end{figure}

\begin{figure}
\begin{center}
\includegraphics[width=10cm]{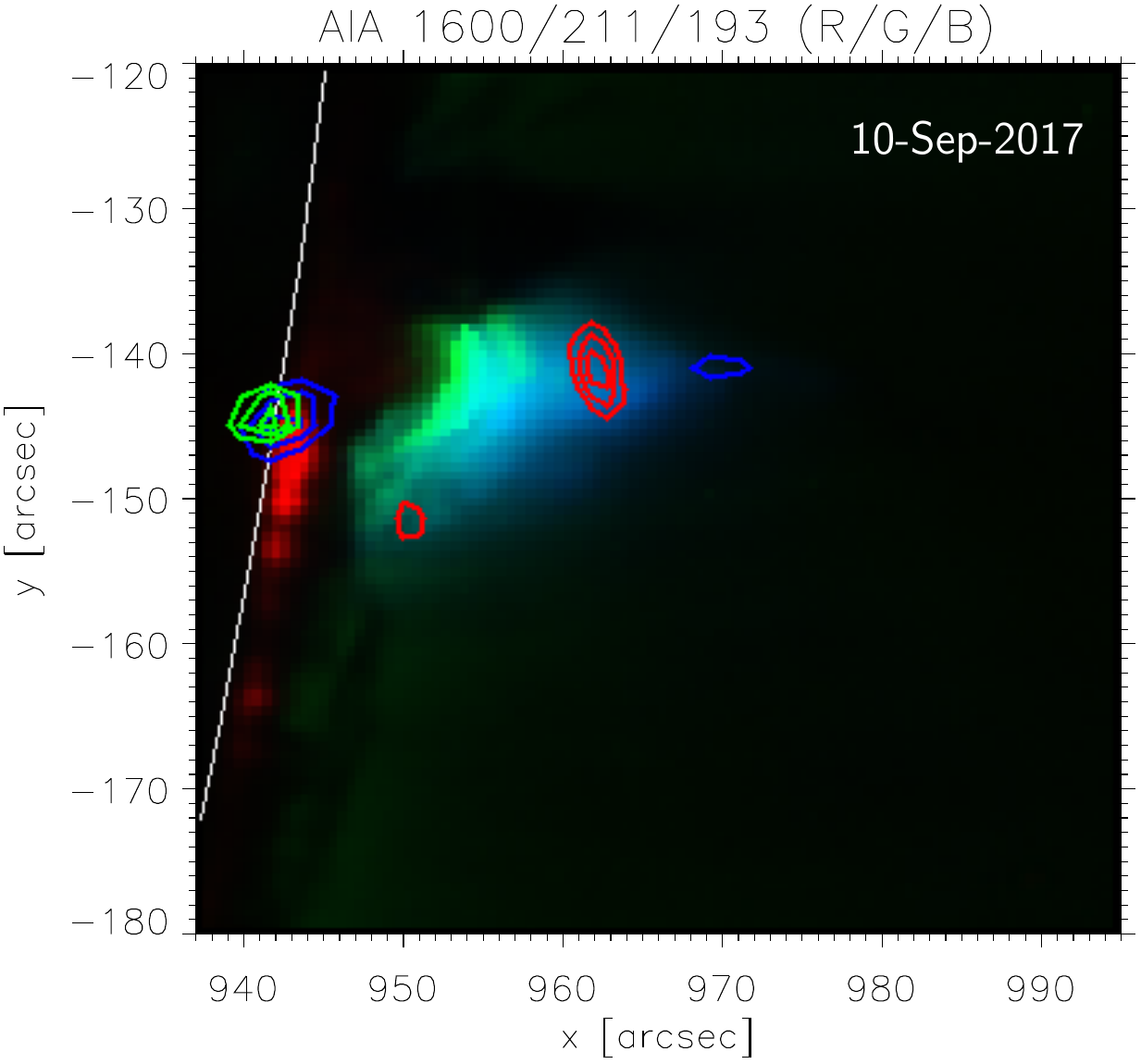}
\end{center}
  \caption{\textit{RHESSI} sources at 15:53 UT during the 2017 September 10 flare superimposed on the 1600~\AA~(red), 211~\AA~(green), and 193~\AA~(blue) \textit{SDO}/AIA images: 6--9 keV (red contours, 50$\%$, 70$\%$, 90$\%$), 25--50 keV (blue contours), 50--100 keV (green contours). The presented field of view corresponds roughly to the cyan box B in Figure~\ref{fig_eovsa}(a).}
  \label{fig_X_source}
\end{figure}

This scenario agrees with our unique observation of the pulsations in two
separate bands in the 1000--1300 and 1600--1800 MHz ranges, just at the moment
of tearing of the ejected filament in the region, where in later times a
signature of the flare current sheet appears (Figures~\ref{fig_initial},
~\ref{fig_171} and ~\ref{fig_4849}). We interpret the pulsations at these two
bands as the plasma emission from the flux rope (1000--1300 MHz) and that from
the top of the flare arcade (1600--1800 MHz). Assuming that the radio emission
of these two bands of DPS at about 15:48:20 UT are emitted on the fundamental
frequency, the plasma density in the flux rope at this early flare phase is in
the range of 1.23 $\times$ 10$^{10}$--2.08 $\times$ 10$^{10}$ cm$^{-3}$ and in
the arcade top is 3.16 $\times$ 10$^{10}$--4.00 $\times$ 10$^{10}$ cm$^{-3}$.
Note that these plasma densities are in the interval between the densities
estimated in the flare arcade at 15:59 UT, i.e. at about 10 minutes after DPS,
\citep[(0.9--2.0) $\times$ 10$^{11}$ cm$^{-3}$][]{2018ApJ...864...63P}) and the
plasma density in the current sheet (10$^{10}$ cm$^{-3}$
\citep{2018ApJ...854..122W}). Because the density in the arcade increases
during the flare and the density in the DPS source is expected to be similar to
that in the current sheet, the estimated densities support our interpretation.
It is interesting that the pulsations in these two bands are interconnected by
fast drifting bursts. We propose that they are generated by superthermal
electrons propagating between the pulsation regions. Moreover, DPS in the
1000--1300 MHz as a whole drifts negatively and the narrowband continuum in the
1600--1800 MHz range, with the shortest period found (0.09--015 s) drifts
positively. It expresses the upward motion of the rising magnetic rope to
lower plasma densities and the increase of the plasma density in the arcade,
respectively, in agreement with our interpretation.

\acknowledgements
We acknowledge support from the project RVO-67985815 and GA
\v{C}R grants 17-16447S, 18-09072S, and 19-09489S. D.G. and B.C. acknowledge support
from NSF grant AST-1910354, AGS-1654382, and NASA grants 80NSSC18K1128,
80NSSC19K0068, and NNX17AB82G to NJIT. This work was also supported by the
Science Grant Agency project VEGA 2/0048/20 (Slovakia). Help from the Bilateral
Mobility Projects SAV-18-01 of the SAS and CAS is acknowledged. This article
was created in the project ITMS No. 26220120029, based on the supporting
operational Research and development program financed from the European
Regional Development Fund. We acknowledge the use of the \textit{Fermi} Solar Flare Observations
facility funded by the \textit{Fermi} GI program (\url{http://hesperia.gsfc.nasa.gov/fermi_solar/}).

\newpage

\end{document}